\documentclass{LT23auth}
\usepackage{graphicx}
\newcommand{\bm}[1]{\mbox{\boldmath$#1$}}

\begin{document}

\begin{frontmatter}
 
\title{Identification of topologically different valence bond states in
 spin ladders}
\author{Masaaki Nakamura}
\address
 {Department of Applied Physics, Faculty of Science,
 Tokyo University of Science, Kagurazaka, Shinjuku-ku, Tokyo 162-8601,
 Japan}
\begin{abstract}
 We discuss relation between three different methods to identify
 topologically distinct short-range valence-bond ground states in
 spin-$\frac12$ two-leg ladders: the string order parameters, the
 level-crossing method, and the ground-state expectation value of the
 twist operator. For this purpose we reexamine a phase transition in the
 frustrated spin ladder.  We propose a proper bosonization of the string
 order parameters. Then, these three methods are shown to be equivalent
 reflecting the $Z_2\times Z_2$ symmetry breaking of the order and the
 disorder fields of the two-dimensional Ising model.
\end{abstract}

%
% write here 3 or 4 keywords separated by semicolons
%
\end{frontmatter}

%%%%%%%%%%%%%%%%%%%%%  Introduction  %%%%%%%%%%%%%%%%%%%%%%%

Since the experimental discoveries of several spin ladder compounds,
there has been considerable interest in properties of various two-leg
spin ladders.  The ground states of these ladders are well described by
short-range valence bond (dimer) states with an energy gap in spin
excitations.  In ladder systems, a variety of topological configurations
of the dimers are possible, so that we expect phase transitions between
topologically different valence bond ground states.

%%%%%%%%%%%%%%%%%%%%%%%%%  Model  %%%%%%%%%%%%%%%%%%%%%%%%%%

As a typical model which exhibits the topological transition, we
consider the antiferromagnetic two-leg spin $S=\frac12$ ladder with
frustration described by
\begin{eqnarray}
 \lefteqn{{\mathcal H}=
  \sum_{j=1}^L ~[J({\bm S}_{1,j}\cdot{\bm S}_{1,j+1}
  +{\bm S}_{2,j}\cdot{\bm S}_{2,j+1})}\\
  &&+J_{\bot}{\bm S}_{1,j}\cdot{\bm S}_{2,j}
   +J_{\times} ({\bm S}_{1,j}\cdot{\bm S}_{2,j+1}
   +{\bm S}_{1,j+1}\cdot{\bm S}_{2,j})],\nonumber
   \label{eqn:model}
\end{eqnarray}
where ${\bm S}_{n,j}$ denotes an $S=\frac12$ operator at site $j$ of the
$n$-th chain.  It has been pointed out that there appears competition
between two phases~\cite{Weihong-K-O,Kim-F-S-S}: For $J_{\bot}\ll
J_{\times}\leq J$, the ground state is similar to that of the $S=1$
Haldane-gap state described by the Affleck-Kennedy-Lieb-Tasaki (AKLT)
model, while for $J_{\times}\ll J_{\bot}$, the system has the same
property of the conventional spin ladder, and the resonating valence
bond (RVB) phase including the rung dimer state is stabilized.

%%%%%%%%%%%%%%%%%%%%%  String order parameter  %%%%%%%%%%%%%%%%%%%%%%%

First, we analyze the phase transition by using the string order
parameter (SOP) which is known to detect the hidden antiferromagnetic
order in the Haldane gap phase of the $S=1$ Heisenberg chain. In ladder
systems, the AKLT and the RVB phases are identified by the following two
SOPs\cite{Kim-F-S-S},
\begin{eqnarray}
 {\mathcal O}^\alpha_{p}=-\lim_{|k-l|\to\infty}
  \left\langle{\tilde S}_{p,k}^\alpha
   \exp\left[{\rm i}\pi\sum_{j=k+1}^{l-1}{\tilde S}_{p,j}^\alpha\right]
   {\tilde S}_{p,l}^\alpha\right\rangle,
\label{eqn:SOPs}
\end{eqnarray}
where $\langle\cdots\rangle$ means the ground-state expectation value,
and $\alpha=x,y,z$. The composite spin operators for $p=$ odd, even are
defined by
\begin{eqnarray}
{\tilde S}_{{\rm odd},j}^\alpha=S_{1,j}^\alpha+S_{2,j}^\alpha,\quad
{\tilde S}_{{\rm even},j}^\alpha=S_{1,j}^\alpha+S_{2,j+1}^\alpha.
\label{eqn:composite-spins}
\end{eqnarray}
Then the AKLT and the RVB phases are characterized by ${\mathcal
O}^\alpha_{\rm odd}\neq0$, ${\mathcal O}^\alpha_{\rm even}= 0$, and by
${\mathcal O}^\alpha_{\rm odd}=0$, ${\mathcal O}^\alpha_{\rm even} \neq
0$, respectively.

The SOPs can be bosonized by using the relation of the $S=1/2$ operator
$\exp({\rm i}\pi S^z_{n,j})=2{\rm i}S^z_{n,j}$.  In the continuum limit,
however, ${\mathcal O}_{\rm odd}^z$ and ${\mathcal O}_{\rm even}^z$ have
the same bosonized form\cite{Kim-F-S-S}: $\lim_{|x-y|\rightarrow\infty}
\langle {\rm e}^{{\rm i}\sqrt{2}\phi_{+}(x)} {\rm e}^{-{\rm
i}\sqrt{2}\phi_{+}(y)}\rangle$ where $\phi_{\pm}$ is the
symmetric/antisymmetric boson field.

%In order to recover information to
%distinguish two SOPs, we consider the following operator,
%%defined in open systems,
%\begin{equation}
% {\mathcal O}_k^z=
%  \exp\left[{\rm i}\pi\sum_{j=1}^{k-1}
%	      \tilde{S}^z_{{\rm odd},j}\right]
%  \tilde{S}^z_{{\rm odd},k}.
%\end{equation}
%One can show that $\langle{\mathcal O}_k^z\rangle$ is pure imaginary in
%the AKLT state.

In order to distinguish two SOPs, we consider the boson representations
of the composite spin operators,
\begin{eqnarray}
  \tilde{S}^z_{{\rm odd},j}&\rightarrow&
  \frac{\sqrt{2}}{\pi}\partial_x\phi_{+}
  +\frac{2{\rm e}^{{\rm i}\pi x/a_0}}{\pi a_0}
  \sin\sqrt{2}\phi_{+}\cdot\cos\sqrt{2}\phi_{-},
  \label{eqn:S_odd_boson}\\
  \tilde{S}^z_{{\rm even},j}&\rightarrow&
  \frac{\sqrt{2}}{\pi}\partial_x\phi_{+}
  +\frac{2{\rm e}^{{\rm i}\pi x/a_0}}{\pi a_0}
  \cos\sqrt{2}\phi_{+}\cdot\sin\sqrt{2}\phi_{-}.
  \label{eqn:S_even_boson}
\end{eqnarray}
Here $\tilde{S}^z_{{\rm odd},j}$ and $\tilde{S}^z_{{\rm even},j}$ are
related by the shift of the phase fields,
\begin{equation}
 \phi_{\pm}\rightarrow\phi_{\pm}\pm\pi/\sqrt{8}.
  \label{eqn:shift}
\end{equation}
This relation indicates that two SOPs are identified as
\begin{eqnarray}
  {\mathcal{O}}^{z}_{{\rm odd}}&\sim&\lim_{|x-y|\rightarrow\infty}
   \langle\sin[\sqrt{2}\phi_{+}(x)]\sin[\sqrt{2}\phi_{+}(y)]\rangle,
   \label{eqn:boson_odd}\\
  {\mathcal{O}}^{z}_{{\rm even}}&\sim&\lim_{|x-y|\rightarrow\infty}
   \langle\cos[\sqrt{2}\phi_{+}(x)]\cos[\sqrt{2}\phi_{+}(y)]\rangle.
   \label{eqn:boson_even}
\end{eqnarray}
Note that Eq.~(\ref{eqn:shift}) reverses the sign of the nonlinear term
of the symmetric mode of the sine-Gordon model.  Thus it turns out that
${\mathcal O}_{\rm odd}^z={\mathcal O}_{\rm even}^z$ gives the critical
point. Besides, since the phase field is related to the order
($\sigma_i$) and disorder ($\mu_i$) parameter of the two dimensional
noncritical Ising models as $\exp({\rm i}\sqrt{2}\phi_+)\sim\mu_1
\mu_2+{\rm i}\sigma_1 \sigma_2$\cite{Shelton-N-T}, two SOPs are
identified as ${\mathcal{O}}^{z}_{{\rm
odd}}\sim\langle\sigma_1\rangle^2\langle\sigma_2\rangle^2$,
${\mathcal{O}}^{z}_{{\rm
even}}\sim\langle\mu_1\rangle^2\langle\mu_2\rangle^2$.  Thus
Eq.~(\ref{eqn:shift}) for $\phi_+$ field corresponds to the
Kramers-Wannier duality transformation $\sigma_i\leftrightarrow\mu_i$
for two Ising copies, and ${\mathcal O}^\alpha_{\rm odd}\neq0$
(${\mathcal O}^\alpha_{\rm even}\neq0$) means the breakdown of the
$Z_2\times Z_2$ symmetry in the order (disorder) fields.

%%%%%%%%%%%%%%%%%%%%%  Level-crossing method  %%%%%%%%%%%%%%%%%%%%%%%

On the other hand, it has been pointed out that a critical point between
two non-degenerate ground states such as the AKLT and the RVB phases can
be identified by the level-crossing of excitation spectra with different
parities under antiperiodic boundary conditions\cite{Kitazawa}.
According to the conformal field theory, operators appearing in the
correlation functions ${\mathcal O}_j$ are related to the excitation
spectra $\Delta E_j$ through the scaling dimensions $x_j$:
\begin{equation}
 \Delta E_j=\frac{2\pi v}{L}x_j,\ \ \ 
  \langle {\mathcal O}_j(x){\mathcal O}_j(y)\rangle\sim |x-y|^{-2x_j},
\end{equation}
where $v$ is the spin wave velocity.  The operators
$\sin(\sqrt{2}\phi_{+})$, $\cos(\sqrt{2}\phi_{+})$ appearing in
Eqs.~(\ref{eqn:boson_odd}) and (\ref{eqn:boson_even}) are nothing but
those corresponding to the energy spectra that give the level-crossing
point.

%%%%%%%%%%%%%%%%%%%%%%%%%%%%  z_L  %%%%%%%%%%%%%%%%%%%%%%%%%%%%%

Finally, we discuss another order parameter.  Recently, the author with
Todo proposed the ground-state expectation value of the twist operator
as an order parameter to characterize valence bond solid states in
quantum spin chains\cite{Nakamura-T},
\begin{equation}
 z_L=\left\langle\exp\left[{\rm i}
      \frac{2\pi}{L}\sum_{j=1}^L j\tilde{S}_{{\rm odd},j}^z\right]
     \right\rangle.
 \label{eqn:def_z}
\end{equation}
The asymptotic form of $z_L$ is given by $z_L=(-1)^n[1-{\mathcal
O}(1/L)]$, where $n$ is the number of valence bonds at boundary.  A
similar relation was also found in two-dimensional dimer
systems\cite{Bonesteel}. Therefore, we obtain $z_L>0$ ($z_L<0$) for the
RVB (AKLT) state.  It has been proposed that $z_L$ defined by
Eq.~(\ref{eqn:def_z}) is identified by the ground-state expectation
value of the nonlinear term of the sine-Gordon
model\cite{Nakamura-V,Nakamura-T},
\begin{equation}
 z_L\propto\langle\cos(\sqrt{8}\phi_+)\rangle\rightarrow
  \langle\mu_1\rangle^2\langle\mu_2\rangle^2
  -\langle\sigma_1\rangle^2\langle\sigma_2\rangle^2.
  \label{eqn:ident_z_L}
\end{equation}
Thus it turns out that $z_L$ has information of two SOPs, characterizing
the $Z_2\times Z_2$ symmetry breaking in the order (disorder) fields by
alternating its sign.  In case that $\tilde{S}_{{\rm odd},j}^z$ in $z_L$
is replaced by $\tilde{S}_{{\rm even},j}^z$, the sign of $z_L$ is
reversed\cite{Nakamura-T2} due to the relation of Eq.~(\ref{eqn:shift}).

From the above argument, three different methods to identify the
topologically distinct valence bond ground states have been shown to be
equivalent. We have also performed numerical analysis for
Eq.~(\ref{eqn:model}) using the exact diagonalization, and observed that
phase transition point determined by these three methods show reasonable
agreement. These three different methods are also applied to the
analysis of phase transitions in the $S=1$ spin ladder with bond
alternation\cite{Yamamoto-S-K,Matsumoto-T-N-Y-T}.

This work is partly supported by the Grant-in-Aid for scientific
research of the Ministry of Education, Science, Sports and Culture of
Japan.

\end{document}